# Privacy-Respecting Type Error Telemetry at Scale

Ben Greenman[a,b], Alan Jeffrey[c], Shriram Krishnamurthi[a], and Mitesh Shah[c]

a   Brown University, Providence, RI, USA
b   University of Utah, Salt Lake City, UT, USA
c   Roblox, San Mateo, CA, USA

**Abstract**
**Context**   Roblox Studio lets millions of creators build interactive experiences by programming in a variant of Lua called Luau. The creators form a broad group, ranging from novices writing their first script to professional developers; thus, Luau must support a wide audience. As part of its efforts to support all kinds of programmers, Luau includes an optional, gradual type system and goes to great lengths to minimize false positive errors.
**Inquiry**   Since Luau is currently being used by many creators, we want to collect data to improve the language and, in particular, the type system. The standard way to collect data is to deploy client-side telemetry; however, we cannot scrape personal data or proprietary information, which means we cannot collect source code fragments, error messages, or even filepaths. The research questions are thus about how to conduct telemetry that is not invasive and obtain insights from it about type errors.
**Approach**   We designed and implemented a pseudonymized, randomly-sampling telemetry system for Luau. Telemetry records include a timestamp, a session id, a reason for sending, and a numeric summary of the most recent type analyses. This information lets us study type errors over time without revealing private data. We deployed the system in Roblox Studio during Spring 2023 and collected over 1.5 million telemetry records from over 340,000 sessions.
**Knowledge**   We present several findings about Luau, all of which suggest that telemetry is an effective way to study type error pragmatics. One of the less-surprising findings is that opt-in gradual types are unpopular: there is an 100x gap between the number of untyped Luau sessions and the number of typed ones. One surprise is that the strict mode for type analysis is overly conservative about interactions with data assets. A reassuring finding is that type analysis rarely hits its internal limits on problem size.
**Grounding**   Our findings are supported by a dataset of over 1.5 million telemetry records. The data and scripts for analyzing it are available in an artifact.
**Importance**   Beyond the immediate benefits to Luau, our findings about types and type errors have implications for adoption and ergonomics in other gradual languages such as TypeScript, Elixir, and Typed Racket. Our telemetry design is of broad interest, as it reports on type errors without revealing sensitive information.

The Art, Science, and Engineering of Programming

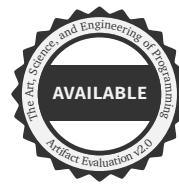
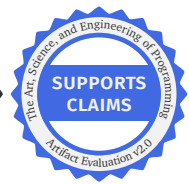

**Privacy-Respecting Type Error Telemetry at Scale**

## 1 Introduction

Roblox is a platform for shared virtual experiences (typically 3D games), with 70 million Daily Active Users, and 16 billion hours of engagement in July–September 2023 [57]. There are over 5 million distinct user-created programs available on the platform thanks to a worldwide community of 3 million creators. Many of the creators program for fun or learning and may not consider themselves software developers; others are professional developers who work full time with the Roblox platform.

Creators program using the Luau programming language [40], an extension of Lua 5.1 [28]. The main addition in Luau is a static type system that infers types for all Luau programs on the fly, as creators modify the code. These types are used primarily in IDE tooling such as autocomplete and in API documentation [58], and indeed, creators may be unaware that a typechecker is analyzing their code. However, creators can opt in to receiving type error reports and they can write their own types to guide designs and document their intentions.

Due to the broad community of creators, the goals of the Luau type system are rather unique [7]. Whereas a traditional type system focuses on compilation and memory safety, Luau takes a lenient approach by default and lets creators gradually [61, 65] migrate to rigorous checks one module at a time via three analysis modes. The default nocheck mode reports only syntax errors, the nonstrict mode reports would-be runtime errors and skips over other issues, and the strict mode reports a variety of potential errors. In addition, no matter the mode, a background type analysis runs to guide IDE tools (Section 2). Luau's types aim to support untyped designs, in the spirit of migratory typing [66], so that creators can switch modes without needing to restructure their code in major, potentially-breaking, ways.

In this paper, we investigate methods for measuring the effectiveness of the Luau type system. The goal is to collect feedback at a large scale, with thousands of participants maintaining real codebases. Consequently, the measurements cannot reveal *any* information about source code, as it may contain personal data, proprietary algorithms, novel game designs, and so on. In comparison to prior work (Section 6), which with few exceptions is small in scale or collects source code, we performed a large-scale study using pseudonymized telemetry.

Our starting point is a telemetry framework that is built in to Roblox Studio and currently measures the effectiveness of creation features. This system randomly determines which sessions should report telemetry, and, for those sessions, reports telemetry records back with a summary of the session.

In this work, we design telemetry that collects data on type errors without exposing source code, source locations, or even error message text (which may contain revealing information). The telemetry counts the number and *kind* of type errors at various levels of granularity. Furthermore, it maintains a client-side approximation of the latest source-code edits and uses that to identify type errors that overlap with this edit range. Each record contains a pseudonymized session identifier and a timestamp.

With this telemetry data, we investigate research questions about the adoption and benefits of type analysis:

**RQ1.** How many sessions use type analysis? How often do projects contain modules with different analysis modes? How often do sessions turn analysis off?

**RQ2.** For modules that use type analysis: which errors arise, how do sessions respond, and which errors tend to persist despite subsequent edits?

**RQ3.** What impact does type analysis have on the number of background errors? For example, do background errors pile up in unanalyzed (nocheck) projects?

Beyond their immediate revelance to Luau, answers to these questions have broad implications for the design of gradually-typed languages. Luau represents a large-scale combination of ideas from gradual typing [5, 61, 66], success typing [38], and semantic subtyping [11, 29]. Lessons from this experience can inform future applications.

At a higher level, this paper is the first to use telemetry to study a type checker. It thus represents a step toward data-driven language design, informed by many users' actual practice. Our data captures over 340,000 sessions that occured between February and April 2023 and covers thousands of type analysis errors and millions of background errors. By contrast to typical qualitative methods such as surveys and interviews, it is not restricted to users' *perceptions* about their work and it is not limited to a small number of users performing synthetic tasks.

**Contributions**

- Design of a low-overhead telemetry system that reports on type errors without revealing potentially private information.
- Lessons from many thousands of type errors about adoption, persistent errors, and creators' responses.
- A dataset of over 1.5 million telemetry records and scripts to analyze them [23].

## 2   Context: Roblox and Luau

Roblox Studio is a workbench that combines 3D creation tools and an Integrated Developer Environment (IDE), as seen in Figure 1. The IDE includes an optional *Script Analysis* widget that reports syntax errors, type errors, and problems identified by lint tools. The main editor widget can also highlight source locations in code where reported errors occur.

**Type Analysis Modes, Background Analysis**   Creators can opt in to detailed error reports and highlights by selecting a *type analysis mode* for each script. There are three modes that creators may choose from:
- nocheck: report only syntax errors (the default),
- nonstrict: report syntax errors and a subset of high confidence type errors, and
- strict: report syntax errors and all type errors.

Each run of the type analysis can report several errors. There is no guarantee that creators read every error. In fact, creators who close the Script Analysis widget can see highlights in their code but no further details about the errors.

**Privacy-Respecting Type Error Telemetry at Scale**

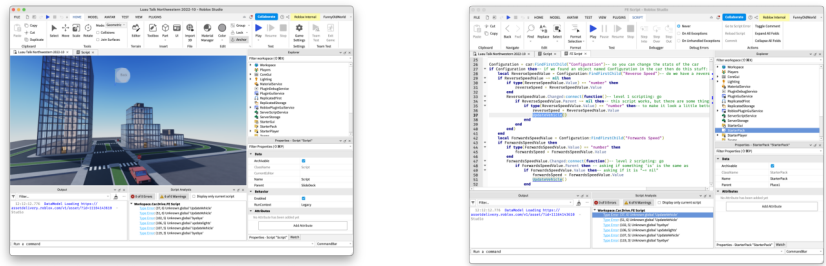

**Figure 1** Roblox Studio 3D creation tools (left) and IDE (right).

As an example of nonstrict mode, the following program reports only one error:

```
--!nonstrict
local x = { p = 5, q = nil }
if condition then x.q = 7 end
local y = x.p + x.q  --> OK
local z = x.r        --> UnknownProperty: Key 'r' not found in table 'x'
```

In strict mode, it reports two errors:

```
--!strict
local x = { p = 5, q = nil }
if condition then x.q = 7 end
local y = x.p + x.q  --> TypeMismatch: Type 'nil' could not be converted into 'number'
local z = x.r        --> UnknownProperty: Key 'r' not found in table 'x'
```

In programs like the above, where it is undecidable whether there will be a run-time error, strict mode errs on the side of reporting an error and nonstrict mode errs on the side of suppressing the error.

Both modes report the UnknownProperty error because misspellings of property names are common enough to report in both strict and nonstrict mode. The nocheck mode reports no error because the program is syntactically valid. See [7] for a more detailed discussion of the rationale behind the type systems.

In addition to the main type analysis, Roblox Studio takes a second pass over every codebase with a *background* analysis to infer autocomplete suggestions and drive other IDE tools. To infer precise types, the background analysis uses rigorous checks very similar to those of strict mode (but not identical: see Section 4.8) no matter what mode the creator declared for the script. When it fails to infer a type, background analysis defaults to the unknown gradual type [61]. Creators cannot see the errors reported by background analysis. Internally to Roblox Studio, this analysis is called *forced strict* because it ignores the creator's choice and applies strong checks.

Table 1 lists a few of the type errors that script analysis can report (10 out of 35 total) to give a sense of Luau. A SyntaxError is the only error that can appear in nocheck code. Several errors including UnknownProperty are about tables, which are Luau's (and Lua's) primary data structure. Tables encode arrays, dictionaries, and objects; consequently, tables can have methods and can usually be extended. The error TypesAreUnrelated is a special kind of type mismatch that arises only during cast,

■ **Table 1** Sample type errors.

| Label | Interpretation |
| --- | --- |
| TypeMismatch | Basic type error. |
| SyntaxError | Basic parse error, e.g., for if end. |
| UnknownProperty | Referenced an invalid field or method. |
| OnlyTablesCanHaveMethods | Tried to attach a method to a non-table. |
| CannotExtendTable | Tried to extend a sealed table. |
| TypesAreUnrelated | Failed cast, unify, or subtype. |
| CountMismatch | Arity mismatch for a function. |
| IncorrectGenericParamCount | Arity mismatch for a generic type. |
| CodeTooComplex | Type analysis failed (Section 4.6). |
| GenericError | Generic label for other non-type errors, e.g., looping over an unordered table. |

unification, or subtyping check. A CountMismatch occurs when the arguments to a function do not match the function's arity. Similarly, an IncorrectGenericParamCount is an arity mismatch for a generic type. A CodeTooComplex occurs when the typechecker hits an internal limit on problem size. Lastly, GenericError is a catch-all label for miscellaneous errors. When Luau can provide more context regarding a generic error, it uses a second error label called ExtraInformation to attach a brief description.

**On-the-Fly Typechecking**   Luau modules provide reusable libraries, which may be imported by other scripts. Imports therefore form a graph in general, though type analysis removes edges to ensure an acyclic graph. In strict mode, it is an error to create an import cycle.

Typechecking analyzes the entire module graph, but it also keeps track of which modules have been edited to avoid repeating work. Any script that the creator modifies gets marked as dirty, and any script that is dirty or that transitively imports a dirty module gets analyzed by the typechecker. When typechecking is performed for autocomplete, the common case is that only the current script gets checked because it is the only dirty script. (Since the module graph is acyclic, nothing that the current script requires can transitively require it.) With this strategy, the typechecker can run after every keystroke without slowing down Roblox Studio.

**Data Model**   The state of the world in a Roblox program is captured by the *data model*, which is a tree of instances, such as parts, models, meshes, effects, lighting, audio assets, and physics constraints such as forces, springs and joints. Each asset lives in a separate file. There may be thousands of assets.

While a program is under development, it is typical for the data model to be edited. Since the shape of the data model tree is reflected in the type system, it is possible for these edits to introduce a huge number of type errors across the project.

## 3 Telemetry Design

Telemetry allows an application to phone home with data summarizing usage patterns, such performance data, crash reporting, and feature uptake. A typical use-case is in deciding whether an API can be deprecated; without telemetry, it may be difficult to know how popular the API is, but with telemetry it is straightforward. Telemetry for programming languages can, however, be controversial. See, for example, the lively discussion around telemetry in the Go toolchain [15].

Roblox Studio has a framework for reporting telemetry. Various features in the IDE use telemetry to measure their effectiveness. The Luau open source toolchain does *not* report telemetry, as it is designed to be used in build environments such as Continuous Integration servers, where hermetic deterministic builds are expected.

We added a subsystem to Roblox Studio's telemetry in order to study type errors in Luau code. This section explains the constraints our telemetry had to meet (Section 3.1) and the final design (Section 3.2).

### 3.1 Limitations for Luau Telemetry

Telemetry in Roblox Studio faces three major constraints:

1. It *must not reveal private information*. This can include Personally Identifying Information (PII) about creators: their identity, location, etc. It also includes trade secrets. Even an error message that contains the name of a function can reveal something the creator does not want to share.
2. On the client side, telemetry computations *must run quickly enough* to avoid slowing down the editing experience.
3. Telemetry *must transmit a small amount of data* to avoid overloading the servers. This constrains both the number of records the system can transmit and the size of each record.

See Cox [16] for further discussion of the privacy implications and tradeoffs of telemetry for programming languages.

In addition, due to the architecture of Roblox Studio, there is some information that is not available to the typechecker:

- Lifecycle events including save, run, quit, and publish. (Publish events cannot be reported at any rate because they are public, and could be used to match a telemetry session to a creator.)
- GUI state: in particular, whether the Script Analysis widget is visible.

Our type error telemetry has to meet all of the above constraints. This means that it cannot send event-based telemetry on IDE lifecycle events (due to Studio's architecture) and it cannot send full results from every type analysis because that would overwhelm the data-collection server.

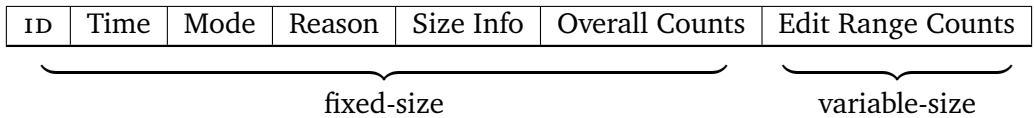

**Figure 2** Structure of a telemetry record. There are 13 overall counts and up to 70 edit range counts in each record.

## 3.2 Telemetry Records

The type error telemetry for a Roblox Studio session is gathered as a series of records, all with the same pseudonymized session identifier. This allows us to correlate telemetry across a single session, but not between sessions. In order to avoid swamping the telemetry servers, the client reports on a subset of telemetry events via two levels of uniform random sampling:

- 1 % of Roblox Studio sessions generate *any* type error telemetry, and
- 0.5 % of all type analyses (approximately 1 in 200 keystrokes) in an enrolled session generate a telemetry record.

In addition to sending telemetry for randomly-selected keystrokes, our system sends a telemetry record every time the session changes focus from one module to another. These records provide important context for mode switches: if one record uses strict mode but the next uses nocheck, then it is critical to know whether the creator downgraded modes or simply switched to another module.

Figure 2 shows the structure of telemetry records:

1. ID: a pseudonymized identifier (15-digit random number) for the current session.
2. Time: one timestamp from the client and one from the server.
3. Mode: type analysis mode of the current file: nocheck, nonstrict, or strict.
4. Reason: a flag that explains whether this record was sent due to a randomly-selected keystroke or a module switch.
5. Size Info: number of lines in the codebase and number of lines in the edit range.
6. Overall Counts: summary of type errors and background errors during the last two invocations of type analysis, and the number of times type analysis hit an internal limit on problem size (see Section 4.6 for details). For each invocation and each kind of analysis error, there are three summary counts:
   - total number of errors,
   - errors in the current module, and
   - errors in the current edit range.
7. Edit Range Counts: list of specific errors that arose in the last two type analyses and overlap with the current edit range. Since there are 35 possible errors (Section 2), there can be up to 70 counts in a telemetry record. Exactly how to interpret this data depends on which analysis it came from:
   - Errors in the latest type analysis are currently visible to the creator. These may have been introduced by the changes in the edit range.

- Errors from the previous type analysis were visible before the creator made the latest edits. These may have motivated some of the changes in the edit range.

To track the edit range, we record a start and end position, which we update appropriately on every edit. This can result in very large and imprecise edit ranges, for example, if the user edits at the beginning and end of the file. The upshot of this strategy is that it reduces the size and complexity of telemetry records because there is only one interval for errors to overlap with.

This telemetry design is admittedly coarse-grained. For instance, it does not distinguish between edits that ignore a type error from edits that remove one error while introducing another. We acknowledge this and other threats in Section 5. The main advantages of this telemetry are the small size of each record and the complete lack of private information. Furthermore, despite the limitations, this telemetry supports a variety of inferences that we showcase in the next section.

## 4   The Data

We collected type error telemetry in Roblox Studio between February and April 2023. Every Studio session had a small random chance of generating telemetry (Section 3.2). The chosen sessions generated a record whenever the user switched modules and randomly on each keystroke. In total, we collected over 1.5 million telemetry records. Roughly two thirds are due to keystrokes.

Figure 3 provides a time-ordered distribution of the data. The x-axis is labeled with a California time zone; the thin vertical lines mark the start of a new day. For each of the 24 hours in each day, there is one blue bar indicating the number of records generated during that hour. Each hour has a few dozen to over 3,000 records, typically about 1,000. We conjecture the very low bars are due to outage on the logging server and that the very tall bars are due to user scripts that generated a huge number of keystrokes. Telemetry ended in early April; however, there are short bars in the mid-April range from developers who waited a few days to update Roblox Studio.

The regular pattern of the peaks suggests that many creators follow a regular schedule; these may be professional developers. The tall weekend (shaded) regions could be because Roblox has a significant school-aged creator community.

### 4.1  Data Cleaning

A close inspection of the data revealed two anomalies:

1. Some records have identical timestamps and session IDs. We keep the first such record and discard the rest. (A human would have to switch modules twice or enter several keystrokes within 1ms to generate two such records legitimately.)
2. A few hundred records (1,533) have negative edit ranges. These are likely due to large deletions. Since the issue affects so few records, we simply ignore their edit ranges. We do, however, use their overall error counts and other uncorrupted fields.

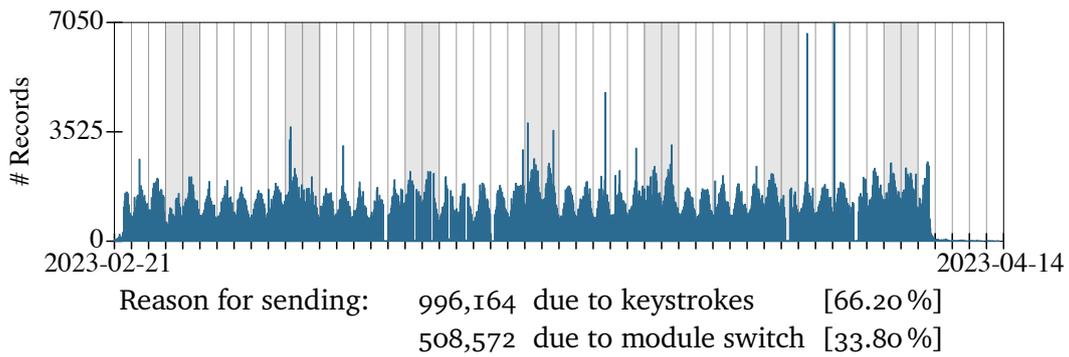

Reason for sending:  996,164   due to keystrokes       [66.20 %]
                     508,572   due to module switch    [33.80 %]

▪ **Figure 3** Telemetry records per hour. Each tick on the *x*-axis marks the start of a new day in California. Shaded ranges correspond to weekends.

▪ **Table 2** Size of analyzed code: number of lines and lines in edit range

|  | Mean | Stddev | Median | P99 | Distribution |
|---|---|---|---|---|---|
| Lines | 6,497 | 22K | 3,115 | 50,547 | |
| Edit Range | 3,680 | 31K | 926 | 34,725 | |

## 4.2 Overall Size and Shape

Three important characteristics of the recorded data are the size of the codebases, the length of the sessions, and the number of type errors. We discuss these in turn.

**Codebase Size** Table 2 summarizes the size of projects in the dataset. The four numeric columns report the mean, standard deviation, median, and 99th percentile for the number of lines and lines in edit range.[1] The plots on the right show zoomed-in distributions of the line and edit range sizes. For example, the x-axis of the first plot counts up to 1,000 lines (not showing a long tail to the right) and the y-axis counts up to 1,000 telemetry records (not showing a tall spike on the left).

There is a huge amount of variation across sessions. The largest ones have over 50,000 lines of code while the smallest have zero lines of code. Unsurprisingly, these wide-ranging numbers come with large standard deviations. The median values are more reasonable, with roughly 3,000 lines of code and 1,000 lines in edit ranges.

---

[1] The data also contains the number of files in each project, but these numbers are difficult to relate to typechecking because projects may contain hundreds of files that define data assets. For the record, the median file count is 7,678 and the 99th percentile is 51,761 files.

◾ **Table 3** Session size in seconds and in number of records.

|  | Mean | Stddev | Median | P99 | Distribution |
|---|---|---|---|---|---|
| Time Span (sec) | 3,184 | 16k | 845 | 35,450 | 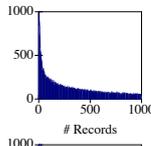 |
| Record Count | 286 | 583 | 138 | 3,302 | 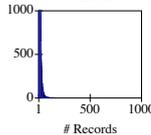 |

◾ **Table 4** Current type errors and background errors across all telemetry records.

| 595,137 | type errors |  | 72,235,735 | background errors |  |
|---|---|---|---|---|---|
| 289,698 | in module | [48.68 %] | 37,027,281 | in module | [51.26 %] |
| 30,924 | in edit range | [5.20 %] | 1,111,178 | in edit range | [1.54 %] |

**Session Size** Table 3 similarly outlines the size of sessions using two metrics: real time in seconds between the first and last record, and the total number of records. The plots focus on the lower-left fragment of the dataset, which again has a very long tail. The longest sessions last several days and/or contain thousands of records, while the shortest last a few milliseconds or include just one record. A median session runs on the order of minutes and consists of a few dozen events.

**Number of Type Analysis Errors** Table 4 lists the number of type analysis errors and breaks them down by location in the codebase. In total, there are over 590,000 type errors. The number of background errors is much larger, at 72 million, because this analysis uses rigorous checks and runs on every module no matter what type analysis mode the module declares. Furthermore, background checks run silently, so creators have no awareness of or incentive to fix these errors. Approximately half the analysis errors occurred in the current module, and a small fraction of these overlapped with the current edit range (5 % for type errors, 1 % for background). The other errors point to different modules in the codebase.

The fraction of errors in the current module is relatively high because projects contain many modules, all of which might contain latent type errors. This high fraction is encouraging for two reasons. First, it suggests that errors appear locally, as the result of edits to nearby code, rather than as the result of edits to code that lives in another module. Second, it suggests that creators fix errors before switching to another module.

The low fraction of errors in the edit range sends a mixed message:

- On one hand, it may be that edits successfully remove errors. This is true, however, only for type errors and not for background errors. There are 11,479 more type errors from the previous analysis that overlap with the current edit range (27 % difference), but 172,678 fewer overlapping previous background errors (15 % difference). Since creators can see type errors but not background errors before making

|           |          |          |         |            |          |
|----------:|----------|----------|--------:|------------|----------|
| 1,341,348 | nocheck  | [89.14%] | 313,509 | nocheck    | [90.19%] |
|   156,883 | nonstrict| [10.43%] |  32,902 | nonstrict  | [9.47%]  |
|     6,505 | strict   | [0.43%]  |     545 | strict     | [0.16%]  |
|           |          |          |     642 | mixed-mode | [0.18%]  |

**(a)** Analysis modes in 1,504,736 records.   **(b)** Analysis mode(s) in 347,598 sessions.

Among the other sessions: 176 contain a mode upgrade, 233 contain a mode downgrade, and 512 contain modules with different modes.

■ **Figure 4** Overview of type analysis modes.

edits, we conclude that visibility makes a difference. Section 4.5 explores edits and type errors in more detail.

- On the other hand, the low overlap between errors (whether current or previous) and edits says that edits rarely target highlighted code. There are several possible explanations: analysis might report several errors and the edits only target one or two of them; creators might ignore errors entirely and edit other code; or creators might choose to edit a non-highlighted location to fix an error. While non-privacy-protecting analyses have answered such questions in the past [46], our data cannot narrow down an answer.

### 4.3 Type Analysis Modes

Creators have three analysis modes to choose from and can switch between modes at any time. Figure 4 shows, however, that usage is extremely skewed toward the default nocheck mode. Nearly 90% of all records use nocheck, while 10% use nonstrict and only a tiny fraction (0.4%) use strict mode. Grouping these records into sessions shows that 90% of all sessions use nocheck exclusively, 9% use nonstrict exclusively, and 0.2% use strict exclusively.

These adoption numbers indicate that as long as type checking is opt-in, most creators stay opted out. In the future, Luau plans to make nonstrict the default mode. It will be interesting to revisit mode usage at that point.

Most sessions (99.82%) stick to a single analysis mode. They never change the mode in the current module and never switch focus to a module with a different analysis mode. Among the other sessions, most of those (80%) switch to a module with a different mode, about half contain at least one edit that upgrades to a more strict mode, and about half contain a downgrade to a less-strict mode. There are 263 total upgrades across all sessions and 320 total downgrades.

**Are Upgrades Discouraging?** A possible explanation for the low adoption of nonstrict and strict mode is that upgrading to these modes leads to a large number of type errors. Creators might get discouraged or overwhelmed by a high error count.

The data does not support this explanation. On average, mode-upgrades resulted in 3 additional type errors (stddev 7, median 0). This data excludes module switch records and considers only keystroke events. The worst-case increase was quite high, at 57 type errors, but exceptional. In another exception, upgrading modes removed three type errors, possibly due to other edits being rolled in with the mode change.

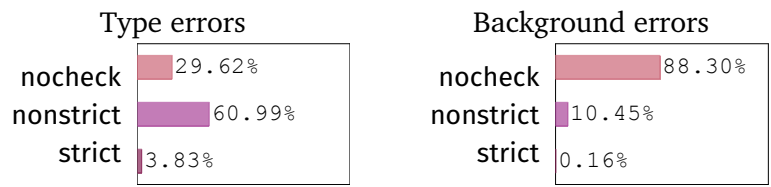

**Figure 5** Type and background errors grouped by mode.

In the other direction, mode downgrades have only a small negative effect on the number of errors (mean -0.3, stddev 4, median 0, max -48). We would see a much larger effect here if creators used downgrades as a quick way to silence the typechecker. But, creators evidently switch modes only when the code is already in a well-typed state.

### 4.4 Errors by Mode

Having seen the total number of errors in the dataset and the popularity of the three analysis modes, a next question is how many errors each mode accounts for. There should be a few type errors in the typical nocheck record (all of which would be syntax errors) and many more in the nonstrict and strict records. For background errors, we would expect very few in strict mode and very many in nocheck mode. Whereas nocheck mode performs little analysis and therefore gives few hints as to how to fix background errors, strict mode checks should closely match background checks.

Figure 5 divides errors across modes. The left plot shows that nonstrict accounts for two thirds of all type errors. This confirms our expectation. The right plot, by contrast, does not support our hypothesis that nocheck records have more background errors than the other modes. Though nocheck does have the most background errors, its *proportion* matches the proportion of nocheck records in the dataset (Figure 4a)—if every record had exactly one background error, the plot would be the same.

Digging further into background errors, the median number of such errors by mode is 1 in nocheck, 3 in nonstrict, and 1 in strict. Records in nocheck mode evidently do not have significantly more background errors. Furthermore, type analysis does not seem to reduce background errors. This is surprising; Section 4.8 explores further.

### 4.5 Type Errors vs. Program Edits

The main focus of our telemetry design is to learn how type errors intersect with edit ranges. Indeed, most of the fields in a telemetry record are dedicated to this topic (Figure 2), and are directed toward RQ2 on type errors.

Using the results of the previous and current type analysis from each telemetry record, Table 5 categorizes errors that overlap with the edit range. The three events of interest are: when edits introduce a type error, when edits preserve a type error, and when edits remove a type error. Hence, the table has three columns for each analysis

▬ **Table 5** Number of telemetry records that increase (↑), preserve (=), or decrease (↓) the amount of a specific type error from the edit range.

|  | nocheck | | | nonstrict | | | strict | | |
|---|---|---|---|---|---|---|---|---|---|
|  | ↑ | = | ↓ | ↑ | = | ↓ | ↑ | = | ↓ |
| CannotCallNonFunction | - | - | - | 12 | 2 | 13 | 1 | - | 1 |
| CannotExtendTable | - | - | - | 6 | 8 | 1 | - | - | 1 |
| CannotInferBinaryOperation | - | - | - | - | 1 | - | 5 | 6 | 6 |
| CountMismatch | - | - | - | 184 | 42 | 157 | 9 | 2 | 9 |
| DuplicateTypeDefinition | - | - | - | 1 | - | - | - | - | - |
| ExtraInformation | - | - | - | 46 | 6 | 33 | 1 | - | 1 |
| FunctionDoesNotTakeSelf | - | - | - | 3 | 4 | 1 | - | - | - |
| FunctionExitsWithoutReturning | - | - | - | 6 | 1 | 5 | 7 | 3 | 5 |
| GenericError | - | - | - | 177 | 47 | 149 | 5 | - | 8 |
| IllegalRequire | - | - | - | 7 | 1 | 11 | - | - | - |
| IncorrectGenericParameterCount | - | - | - | - | - | - | 1 | - | 1 |
| MissingProperties | - | - | - | 8 | 3 | 6 | 5 | 3 | 3 |
| ModuleHasCyclicDependency | - | - | - | 9 | 5 | 8 | - | - | - |
| NotATable | - | - | - | 4 | 1 | 5 | 2 | - | 1 |
| OccursCheckFailed | - | - | - | - | - | - | - | - | 1 |
| OnlyTablesCanHaveMethods | - | - | - | - | - | 2 | - | - | - |
| OptionalValueAccess | - | - | - | 21 | 45 | 15 | 4 | 2 | 4 |
| SyntaxError | - | - | 8290 | - | - | 1149 | - | - | 29 |
| TypeMismatch | - | - | - | 103 | 51 | 80 | 13 | 6 | 18 |
| UnknownPropButFoundLikeProp | - | - | - | 20 | 17 | 13 | - | - | - |
| UnknownProperty | - | - | - | 256 | 156 | 208 | 16 | 18 | 22 |
| UnknownRequire | - | - | - | 43 | 30 | 37 | 5 | 3 | 3 |
| UnknownSymbol | - | - | - | 1992 | 438 | 1797 | 38 | 18 | 35 |

mode. In the nonstrict columns, for example, seventeen records introduce (↑) at least one CannotCallNonFunction error. Four records keep the number for that error at the same, nonzero level, and twenty reduce the number of CannotCallNonFunction errors.

This table is based on the 996,164 records generated from keystroke events. It ignores records based on module switches because those do not have meaningful edit ranges. One implication of the randomized strategy is that we may not notice when a creator fixes an error; the fix must be selected by the telemetry system.

**Observations**

- The numbers in the table are low overall. For instance, the highest strict count is 38 records out of the 6,505 total (Figure 4). Section 4.2 discusses the implications of this low overlap rate for future work.
- SyntaxError and the related errors UnknownSymbol and UnknownProperty are the most common. If the edit range errors are indicative of type errors at large, then

  most errors in the data are likely due to typos during edits. These errors also survive (=) the more than any other.

  Curiously, the number of syntax errors never decreases or stays the same. This may be because the easy fix is to press undo; telemetry does not see undo events.

  The number of typos is inflated by our telemetry strategy because it cannot determine when a creator is mid-edit. For example, a creator who writes a method call bucket.countFish() letter-by-letter will generate an UnknownSymbol error for buck and UnknownProperty for bucket.c, both of which may end up getting sampled. Section 7 suggests ways to avoid reporting typos in the future.

- After the syntax errors, CountMismatch, TypeMismatch, and GenericError are the next most common nonstrict and strict errors. A CountMismatch sends the wrong number of arguments to a function. It may simply be a common error, but it may also be a telemetry artifact. The high incidence of TypeMismatch and GenericError calls for further study in future work to determine precisely which error arose. Some of the GenericErrors come with context strings in the form of ExtraInformation errors, but telemetry does not report these strings.
- OptionalValueAccess frequently persists through edits (=). Creators may be ignoring this error because they can deduce that the optional value will be present at runtime.
- Ideally, the highest number in each column group should be decreases (↓). A decrease should happen when the creator sees an error and fixes it. However, decreases are highest in only 18 of the 39 nonzero groups.

### 4.6 Type Error Frequency

Table 6 sorts type errors by the frequency with which they appear in nonstrict and strict mode. The purpose of this table is to discover and rank all errors that appear in the dataset; thus, it is based on both keystroke and module-switch telemetry records, even though edit ranges are not well-defined for the latter. The table does not include nocheck because all its type errors are SyntaxErrors by definition.

**Observations**
- Typo-level errors, in particular UnknownSymbol (for an unbound variable), are much more common in nonstrict than strict. This may be because nonstrict reports fewer errors to dilute the proportion of unknown symbols.
- Despite being less strict, nonstrict reported four errors that never appeared in strict records: DuplicateTypeDefinition, FunctionDoesNotTakeSelf, MissingUnionProperty, and OnlyTablesCanHaveMethods. We attribute this to the much larger number of nonstrict records in the dataset.
- TypeMismatch is far more common in strict mode (20.46 %) than nonstrict (2.44 %) because strict has additional type constraints. Several other errors have higher strict percentages for the same reason, such as CannotInferBinaryOperation, and two errors appeared only in strict: OccursCheckFailed and IncorrectGenericParamCount.
- Several Luau errors never appeared in edit ranges. Four of these deal with internal typechecker limits, which we discuss below. The others are for dynamic property lookups, deprecated APIs, swapped generic parameters, functions that require a self, and type pack mismatches. We would not expect these to appear often.

**Table 6** Type error popularity for nonstrict and strict modes.

| nonstrict | | strict | |
|---|---|---|---|
| UnknownSymbol | 62.13 % | UnknownSymbol | 23.97 % |
| SyntaxError | 15.42 % | TypeMismatch | 20.46 % |
| UnknownProperty | 8.28 % | UnknownProperty | 18.88 % |
| UnknownRequire | 3.13 % | SyntaxError | 9.31 % |
| TypeMismatch | 2.44 % | CannotInferBinaryOperation | 6.94 % |
| CountMismatch | 2.26 % | MissingProperties | 4.04 % |
| OptionalValueAccess | 2.07 % | CountMismatch | 2.99 % |
| GenericError | 2.06 % | OptionalValueAccess | 2.99 % |
| UnknownPropButGotLikeProp | 0.44 % | FunctionExitsWithoutReturn | 2.99 % |
| CannotExtendTable | 0.42 % | UnknownRequire | 2.37 % |
| GenericExtraInformation | 0.32 % | GenericError | 2.28 % |
| ModuleHasCyclicDependency | 0.23 % | NotATable | 1.32 % |
| IllegalRequire | 0.20 % | GenericExtraInformation | 0.35 % |
| NotATable | 0.15 % | UnknownPropButGotLikeProp | 0.26 % |
| CannotCallNonFunction | 0.15 % | IncorrectGenericParamCount | 0.18 % |
| MissingProperties | 0.09 % | CannotCallNonFunction | 0.18 % |
| FunctionExitsWithoutReturn | 0.07 % | IllegalRequire | 0.18 % |
| FunctionDoesNotTakeSelf | 0.07 % | ModuleHasCyclicDependency | 0.09 % |
| MissingUnionProperty | 0.02 % | CannotExtendTable | 0.09 % |
| CannotInferBinaryOperation | 0.02 % | OccursCheckFailed | 0.09 % |
| OnlyTablesCanHaveMethods | 0.01 % | TypesAreUnrelated | 0.09 % |
| DuplicateTypeDefinition | <0.01 % | | |
| TypesAreUnrelated | <0.01 % | | |

**Internal Limits, Code Too Complex** To deal with pathologies such as the worst-case time for ML-style type inference [33, 45], the Luau typechecker has internal limits that restrict the problems it will attempt to solve. Hitting a limit triggers one of the following errors along with a message that the analysis failed to understand the code: CodeTooComplex, NormalizationTooComplex, or UnificationTooComplex. These errors can appear in Script Analysis just like any other error, but one design goal of Luau is that code rarely hits the limits. A typical creator should never see these errors.

Fortunately, the data rarely contains too-complex errors. None of these errors appear in edit ranges (Table 6). For the specific case of CodeTooComplex, our telemetry tracked project-wide counts and found only 26 errors, which were spread across eleven records in three sessions. Note that all three sessions *could* have been from the same codebase.

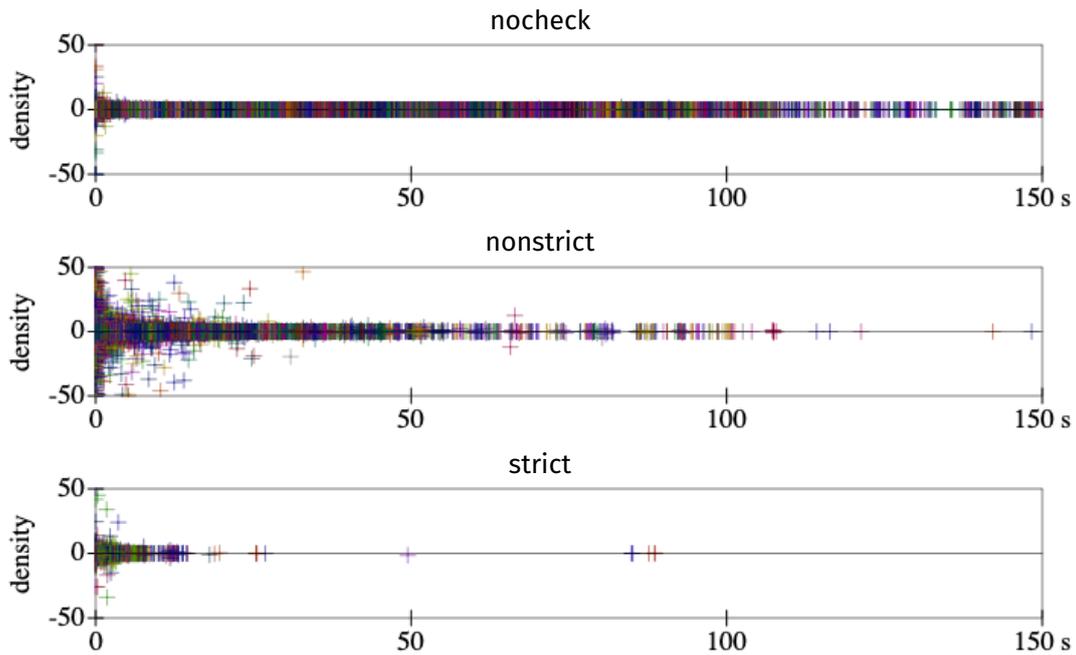

**Figure 6** Changes in type error density ($\frac{\#\text{curr - }\#\text{old errors}}{\#\text{lines}}$) over time (seconds).

### 4.7 Density of Type Errors

Since errors and edits may not overlap, either because the highlights are misleading or because the creator ignores the errors, it is important to see how the total number errors changes over time. We measure *error density* rather than error count, however, to normalize for the wide range of codebase sizes (Table 2). Density is the number of type errors divided by the number of lines.

Figure 6 plots changes to error density for every session in the dataset. The data excludes module-switch telemetry records to keep the focus on edits. The x axis counts time in seconds from the start of the session. Most sessions last a few dozen seconds (Table 3), so the axis ends at 150 seconds to show a bit of the long tail. The y axis shows changes to errors over time; a positive number means errors increased since the last record, and a negative number represents a decrease. While there are a few errors that go outside the plot bounds (max nocheck: 51, max nonstrict: 3586, max strict: 87), we focus on ±50 errors because most of the mass is in that region. The points are color-coded by their session. There are many colors througout the plot, which shows that the data comes from many different sessions rather than from a few high-traffic sessions.

**Observations**
- All three plots are roughly balanced, with equal "mass" above and below the x-axis. The next section explores this point in detail.

|  |  | ↑ | = | ↓ |
|---|---|---|---|---|
| type error (in module) | nocheck | 0 [n/a] | 0 [n/a] | 0 [n/a] |
|  | nonstrict | 7185 [36.61 %] | 4786 [24.30 %] | 7672 [39.09 %] |
|  | strict | 251 [35.35 %] | 204 [28.73 %] | 255 [35.92 %] |
| background error (in module) | nocheck | 153913 [36.96 %] | 152735 [36.68 %] | 109807 [26.37 %] |
|  | nonstrict | 18573 [37.09 %] | 17604 [35.15 %] | 13905 [27.76 %] |
|  | strict | 272 [32.04 %] | 317 [37.34 %] | 260 [30.62 %] |

**Figure 7** How often do edits increase, maintain, or decrease the number of errors?

- Only a small fraction of nocheck records have nonzero type error density. This is in line with Table 5. Module switches in nocheck have nonzero density more often, but still rarely (3 %).
- After the 40-second mark, changes to type error density become much smaller. These long-running sessions are not introducing errors.
- nonstrict has a bigger fluctuations in errors than strict. One explanation is that some nonstrict sessions ignore type analysis (in particular the one with over 3,000 errors!) whereas most strict sessions pay close attention to it.

## 4.8 Do Edits Tend to Add Type Errors?

The visual symmetry in the density plots (Figure 6) suggests that edits add and remove errors with rougly equal frequency, regardless of mode. Figure 7 explores this question of balance in detail. For each of the three modes, it reports the percent of edits that increase (↑), maintain (=), or decrease (↓) the number of type errors in the module. It also reports counts for background checks to see whether fixing type errors fixes background errors as well. The "maintain" rows exclude cases where there are zero errors before and after.

**Observations**
- None of our nocheck samples captured a change in syntax errors in the current module. The nonzero changes in density above (1802 total, shown in Figure 6) all affected other modules.
- The type error increases and decreases are evenly balanced in nonstrict and strict modes (±3 %). Since this data is from keystrokes rather than module switches, it appears that sessions tend to fix errors highlighted by type analysis.
- strict mode has a higher percent of "maintain" (=) records. Errors may be more likely to persist in this mode.
- In background mode, the percent of "maintain" records is higher across the board and roughly equal to the percents of increases and decreases (about one-third each). No matter the mode, reducing type errors is not guaranteed to reduce background errors as well. This is surprising for strict mode, and likely relates to the data model; see below for details.

**Data Model Types: Strict vs. Background Analysis**    Based on the description of background analysis from Section 2, it should raise *at least as many errors* as strict mode because it uses strict checks for all dependencies. For example, if a strict module imports from a nocheck one, then strict analysis does not analyze the import but background analysis does.

In Figure 7, however, nearly 3 % of strict-mode records increase the number of strict errors without increasing background errors. Further inspection reveals that the situation is a bit worse, as some records increase strict errors without changing the background errors: 16 % of all strict records increase (↑) the number of type errors but maintain (=) the number of background errors.

The reason for the discrepancy is that strict mode and background analysis assign different types to the data model: strict assigns a top type and requires downcasts at use sites, whereas background analysis uses the dynamic gradual type. Dealing with these data model errors may be a source of frustration for creators using strict mode.

## 5    Answers to Research Questions

With the data in hand, we can answer the research questions from Section 1:

**RQ1.**
  - **How many sessions use type analysis?**  A mere 10 % of sessions (and of telemetry records overall) opt-in to type checking (Figure 4). Most of these use nonstrict mode, and fewer than 1 % use strict mode.
      Luau has an efficient type checker that can analyze thousands of lines of code (Table 2) quickly enough to be used on-the-fly in an IDE. But, few sessions use type analysis directly. We conjecture that Luau needs further tailoring to match common idioms in untyped code. Discovering these idioms is a question for future work (Section 7).
  - **How often do projects contain modules with different modes?**  Sessions rarely combine analysis modes. Only 512 of 340K sessions (< 0.15 %) switched to modules with different analysis modes. This number may be an underapproximation because we have data only on modules that the creator chose to open, but the widespread use of nocheck mode suggests that it is not off by much. If 90 % of *all* modules indeed use nocheck, then multi-mode projects must be rare.
  - **How often do sessions turn analysis off?**  Sessions rarely disable type analysis after opting in. Only 233 of the 340K sessions (< 0.07 %) switch to a weaker mode. One-fourth of the downgrades switch from strict to nonstrict, and therefore keep some type analysis. The remainder switch from strict to nocheck (50 %) or from nonstrict to nocheck (25 %). Then again, sessions rarely turn analysis on as well: only 176 (< 0.05 %) switch to a stronger mode.

**RQ2. For modules that use type analysis (strict and nonstrict):**
  - **Which errors arise?**  Syntax errors are extremely common (over 50 %, Table 6), followed by type mismatches (20 % in strict), arity errors (2 %), and failure to unpack on optional value (2 %). There are several uncommon errors, such as

- misusing a table operation, and a few that never arise: internal and too-complex errors, duplicating or swapping a generic parameter, use of a deprecated API, and unsafe dynamic property lookup on a class.
- **- How do sessions respond? Which errors persist through edits?** Sessions appear to fix analysis errors. With few exceptions, increases and decreases in the number of errors balance out (Figure 7 and Table 5). Even in nocheck, most edits that overlap with syntax errors remove at least one error (67 %) rather than increase or maintain the total. The errors that survive edits most often are due to inference at a binary operation, sending self to a function that does not expect it, and failure to unpack an optional value. But, the first two seldom arise and the third is something creators might reasonably ignore while prototyping.

**RQ3. What impact does type analysis have on background errors?** Opting in to analysis has little impact on the number of background errors. The percent of background errors in nocheck code matches the percent of sessions using nocheck mode (Figure 5); whereas, if type analysis helped reduce background errors, nocheck would own a higher percentage of the errors. Furthermore, there is no apparent relation between background errors and edits, regardless of analysis mode (Figure 7). A randomly-chosen edit has a roughly 1/3 chance of increasing, decreasing, or maintaining the number of background errors.

**Threats to Validity** While this study has high ecological validity due to its focus on working creators, there are several threats to keep in mind. Some threats stem from our method of collecting data; others come from the limited scope of the data.

First of all, sampling is necessarily incomplete. Though the data includes many thousands of sessions, it may have missed a few critical sessions where a creator adopted strict mode, encountered CodeTooComplex errors, and gave up. Furthermore, our method of sampling by keystroke and by module switch skews the data toward creators who, for whatever reason, do more of these actions. We have no access to other events, e.g., local runs (Section 3.1); future studies may wish to avoid this limitation by actively recruiting participants.

Second, the edit ranges are a coarse approximation of actual edits. They begin at the lowest edited line of code and end at the highest edited line, even if nothing between those lines changed. On a related note, the computation that determines whether an old type error overlaps with the current edit range has many edge cases, especially to handle deletions, and may mis-count. This could explain why Table 5 reports zero syntax errors that overlap with the current edit range; if the fix is to delete code, then the edit range is empty.

Third, session-based data has limitations. A creator who closes and reopens Roblox Studio every day has a much higher chance of generating telemetry than one who leaves Studio open all week. We have no way to reliably check whether multiple sessions are reporting on the same codebase. Sessions may end abruptly, e.g., because of machine reboot, instead of ending when the creator was done editing code.

Fourth, the total counts of type errors include a large number of syntax errors, which have little to do with the Luau type system and are presumably easy for creators to fix. Our comparison between background errors in nocheck vs. strict mode (Figure 7)

◼ **Table 7** Comparison of telemetry designs. Two orthogonal concerns are *when* to collect data and *whether* to add noise to enhance privacy.

|  | Event Counts | Timestamps | Error Msgs. | Source Code |
| --- | :---: | :---: | :---: | :---: |
| Roblox Studio | ● | ● | | |
| Transparent | ● | | | |
| Classic | ● | ● | ● | ● |

might tell a different story without the noise that syntax errors introduce, but we cannot easily filter them because we have only the total number (not the kind) for errors outside the edit range.

Fifth, type analysis reports several errors at once and we have no idea which errors, if any, creators intended to fix with their edits. Thus, fixes may be a side effect of other plans, as seen in prior work [46]. Edits that fix one error but introduce another are also indistinguishable from edits that maintain the number of errors.

Lastly, regarding external validity, we know nothing about the creators in our study except that they were selected at random from a large and diverse group. A dataset based on a targeted subset of users might show entirely different characteristics.

Ultimately, we view our work as primarily *formative*. It gives us a first look at usage, and generates several questions that can lead to hypotheses. Answering those would require much more intrusive techniques, or controlled studies, or other means.

## 6 Related Work

Our telemetry has two distinguishing characteristics: it includes no private information (both PII and source code details), and it sends data on randomly-selected runs of the type checker rather than specific events of interest. To the best of our knowledge, this design occupies a unique position relative to prior work, the two main strands of which we describe below and summarize in Table 7.

In a *classic* telemetry system, such as the ones in VSCode [52], IntelliJ [30], .NET [48] and Eclipse [18], telemetry may include all sorts of data: error reports, source code fragments, timestamps, and filepaths. Users may be able to opt out of some telemetry, but the details depend on the license agreement. Furthermore, at least in VSCode, IDE extensions can report their own telemetry. While this data can be invaluable for discovering bugs in production, it must be handled with extreme care.

At the opposite end of the privacy spectrum, Go's design for *transparent* telemetry reports only counter values [16]. Unlike Roblox Studio, transparent telemetry includes no timestamps and no session IDs. While useful for learning about the frequency and distribution of specific events, the lack of timestamps and IDs means it cannot track edits over time (which we use in Figure 6).

**Telemetry Enhancements: What and How to Sample**  One way to futher strengthen the privacy of any telemetry design, including transparent telemetry, is to systematically add noise to the data using techniques from differential privacy [20, 68, 72]. Recent work shows how to add noise to traces [72], how to account for known relationships between events [73], and how to choose the noise on the client side [25]. Roblox Studio could benefit from these techniques. Applying differential privacy is subtle, however. While we could fuzz the total number of type errors, if we were to fuzz the number of `CodeTooComplex` errors, it might drastically affect our inferences.

Prio and Poplar demonstrate a telemetry method to compute public aggregates over private data using a middle layer of servers [6, 14]. Clients can depend on privacy guarantees even if other clients and some servers (but not all) are malicious. The design has been deployed in Mozilla Firefox to monitor blocklist usage [19]. Luau telemetry might benefit from a redesign to use middle-layer servers, though it is unclear what private data from type analysis would support meaningful aggregates.

Cooperative bug isolation is a method for designing telemetry systems [37]. The goal is to collect a small amount of data from each user and perform statistically-sound analyses; privacy is not a main concern. Each feature of interest *within* a user's codebase has a uniform-random probability of contributing telemetry. A system-builder can analyze this telemetry with predicates to identify notable events and use logistic regression to narrow down nondeterministic bugs. There is a high burden on designers to decide what to capture, but a careful design can minimize the data from each individual user.

Two further refinements, which reduce the burden on experts to select points of interest, are adaptive bug isolation [1] and blame-proportional logging [41]. Adaptive bug isolation starts with a set of predicates, studies which are most correlated with failures, and experiments with adding telemetry to nearby predicates; this technique can reduce the overhead of telemetry by two orders of magnitude relative to naive binary instrumentation. Blame-proportional logging starts with lightweight telemetry to recognize defects, assigns ranks to methods estimating their likelihood as root causes, and uses the ranks and future observations to narrow down the cause. Deep transfer learning is another promising way to hone in on events of interest [74].

**User Studies, Errors, and Type Errors**  Our approach to data analysis is inspired by Marceau et al. [46, 47] and Macedo et al. [43, 44] for Racket and Alloy4Fun, respectively. Other more recent work has also studied edit sequences to infer intent [39, 55, 71]. A major difference is that we have only samples, not full sequences, which significantly impacts our study methods and inferences.

There is a rich body of work that uses IDE instrumentation to study programmers (e.g., [2, 17, 26, 31, 35, 49, 51, 53, 56, 69, 70]). The Blackbox project has accumulated terabytes of Java code over a ten-year span, and has thereby enabled dozens of contributions to CS education [8, 9]. The GRUMPS telemetry system collected over four million keystroke-level actions to study student programmers [62, 63]. WatchDog records the time that programmers spend reading, writing, and testing code [3]. TDD-Guide collects similar data to teach testing methods [50]. None of the above are sufficiently private to work for Roblox Studio. The BitFit project counts notable events (compile program, ask for hint, etc.) for Java [21, 32]. While it does operate in a transparent [16] manner, it provides only coarse-grained information that would be insufficient in our study.

Much work [27, 42] on the benefits of static types focuses on small case studies [24, 64, 67], interviews [12, 13, 22], or static corpuses [4, 54]. Controlled experiments are rare; Kleinschmager et al. [34] is a notable exception. Lerner et al. [36] apply a type error repair tool on over two thousand programs. Seidel et al. [59, 60] use five thousand programs to train data-driven methods for localizing errors. These are all on a vastly smaller scale than our study and also study different issues.

## 7 Lessons Learned and Reflections

This paper presents the first large-scale analysis of developers' interactions with an industrial strength typechecker in a privacy-respecting way. Despite the intentionally-restricted data, the analysis offers several lessons for Luau in particular and for other gradual languages (e.g. [5, 10, 66]) in general:

- Nearly all sessions (90 %) use the default "no check" typechecking mode (Figure 4). Since the next typing mode (nonstrict) requires no annotations and is very conservative about the errors it reports, we conjecture that developers would have no objections to using it and are simply unaware that other modes exist. Luau, and perhaps other languages, must take care to advertise gradual types and demonstrate how to use them effectively to increase adoption.
- Sessions appear to fix errors immediately rather than dealing with them in a batch or ignoring the errors altogether (Figures 6 and 7). This observation should give Roblox confidence to make nonstrict the default in the future. Other languages should consider adding a nonstrict mode that minimizes false positives as a bridge toward strict type checking.
- All Luau scripts interact with data assets, but no typechecking mode has a precise view of the data. Improving precision is an important direction for future work. Until then, strict mode clearly needs to relax its data model analysis because it raises even more errors than background checking (Section 4.8). Other languages should avoid making the same mistake.

On the meta level, the main takeaway is that the lightweight, counter-based telemetry in Roblox Studio supports a variety of useful analyses:

- Counting specific type errors helps to identify trends and to confirm that undesirable events rarely happen (e.g., UnificationTooComplex).
- Testing the relationship between two counts can yield insights. The gap between strict and background errors in Section 4.8 revealed an issue with strict checking.
- Reporting the overlap between type errors and the current (approximate) edit range lets us assess the outcome of edits without revealing source code. It was especially useful to have overlaps for two sets of type errors: old and current.
- Timestamps give useful, low-level insights about the frequency of keystrokes, the length of sessions, and actions over time. The event ordering that timestamps provide was invaluable; e.g., for learning that error deltas bounce up and down rather than rising sharply, then falling (Figure 6).

One aspect of the telemetry that we would change for next time is to provide more metadata. For example, Luau recently adopted semantic subtyping to reduce false-positive type errors [11, 29], but the current telemetry has no way to tell if this internal extension is enabled or not.

Three other aspects are also worth rethinking:

1. It would be useful to track the identity of errors (to know exactly which ones get fixed), to have fine-grained counts for background analysis errors, and to record the specific reason behind errors such as TypeMismatch. But, this additional data could easily double the size of telemetry records.
2. Although tracking edit ranges was useful, it made the telemetry system much more difficult to build and maintain. Tracking old and current errors at the module level would be far easier, though there is a risk that it is too coarse.
3. The massive number of syntax errors begs the question of how to skip them. Shifting focus from arbitrary keystrokes to selected ones, such as closing parentheses or whitespaces, might increase the likelihood of well-formed code without losing the "middle of things" nature of the data. Another option is to ignore records that have a syntax error in the edit range.

It would also be interesting to build statistical models of the "average" programmer using each typing mode and compare their error rates. While one could start modeling with the nocheck and nonstrict data, there are too few strict records at the moment due to our uniform sampling rate. The errors that we do see in strict mode call for an in-depth analysis via talk-aloud interviews to discover why type mismatches occur and how Luau can better accommodate untyped idioms.

**Data Availability Statement**  The artifact for this paper contains the full telemetry dataset and scripts for reproducing the analysis in Section 4.

**Acknowledgements**  Thanks to Benjamin Chung for several helpful discussions about data analysis and effective plotting. We gratefully acknowledge support from NSF grant SHF-2227863 and NSF grant 2030859 to the CRA for the CIFellows project.

**Privacy-Respecting Type Error Telemetry at Scale**

## About the authors

**Ben Greenman** (benjamin.l.greenman@gmail.com) was a CIFellows 2020 postdoc at Brown University and is now an assistant professor in the Kahlert School of Computing at the University of Utah.
https://orcid.org/0000-0001-7078-9287

**Alan Jeffrey** (ajeffrey@roblox.com) is a Principal Software Engineer at Roblox.
https://orcid.org/0000-0001-6342-0318

**Shriram Krishnamurthi** (shriram@brown.edu) is the Vice President of Programming Languages (no, not really) at Brown University.
https://orcid.org/0000-0001-5184-1975

**Mitesh Shah** (mshah@roblox.com) is Senior Engineering Director, Programmability, at Roblox.
https://orcid.org/0009-0000-6084-123X